\documentclass[10pt,conference]{IEEEtran}
\usepackage{amsfonts}
\usepackage{amssymb}
\usepackage{amsmath}
\usepackage{amsthm}
\usepackage{algorithm}
\usepackage{algpseudocode}
\usepackage {graphicx}
\usepackage{bigstrut} 
\usepackage{multirow}
\usepackage{url}
\usepackage{cite}
\usepackage{framed}

\ifCLASSOPTIONcompsoc
\usepackage[caption=false,font=normalsize,labelfon
t=sf,textfont=sf]{subfig}
\else
\usepackage[caption=false,font=footnotesize]{subfig}
\fi

\newtheorem*{mydef}{Definition}
\def\x{{\mathbf x}}
\def\b{{\mathbf b}}

\def\v{{\mathbf v}}
\def\r{{\mathbf r}}

\def\a{{\mathbf a}}
\def\0{{\mathbf 0}}
\def\A{{\mathbf A}}

\def\T{{\mathcal T}}

\def\R{{\mathbb R}}
\def\E{{\mathbb E}}
\def\THat{{\hat\T}}

\newcommand{\vecLTwoNorm}[1]{{\left\lVert#1\right\rVert}_2}
\begin{document}

%
 \title{Fusion of Greedy Pursuits for \\Compressed Sensing Signal Reconstruction}
\author{
\IEEEauthorblockN{Sooraj K. Ambat}
\IEEEauthorblockA{Statistical Signal Processing Laboratory \\
Dept.\ of Electrical Communication Engg.\\
Indian Institute of Science \\
Bangalore, 560012, India.\\
{\tt sooraj@ece.iisc.ernet.in}} \and
\IEEEauthorblockN{Saikat Chatterjee}
\IEEEauthorblockA{Communication Theory Laboratory \\
School of Electrical Engineering \\
KTH Royal Institute of Technology \\
Stockholm, 10044, Sweden.\\
{\tt sach@kth.se}}
\and
\IEEEauthorblockN{K.V.S. Hari}
\IEEEauthorblockA{Statistical Signal Processing Laboratory \\
Dept.\ of Electrical Communication Engg.\\
Indian Institute of Science \\
Bangalore, 560012, India.\\
{\tt hari@ece.iisc.ernet.in}}
}
\maketitle

\begin{abstract}
 Greedy Pursuits are very popular in Compressed Sensing for sparse signal recovery. Though many of the Greedy Pursuits possess elegant
theoretical guarantees for performance, it is well known that their performance depends on the statistical distribution of the non-zero
elements in the sparse signal. In practice, the distribution of the sparse signal may not be known {\em a priori}. It is also observed that
performance of Greedy Pursuits degrades as the number of available measurements decreases from a threshold value which is method
dependent. To improve the performance in these situations, we introduce a novel fusion framework for Greedy Pursuits and also propose two
algorithms for sparse recovery. Through Monte Carlo simulations we show that the proposed schemes improve sparse signal recovery in clean as
well as noisy measurement cases.
\end{abstract}

\section{Introduction}
{\let\thefootnote\relax\footnotetext{$Notations:$ Bold upper case and bold lower case Roman letters denote matrices and vectors
respectively. Calligraphic letters and Upper case Greek alphabets are used to denote sets. $\|.\|_p$ represents the $p^{\mathrm{th}}$
-norm. $\A_\T$ denote the column sub-matrix of $\A$ formed by the columns of $\A$ listed in the set $\T$.
 $\x_\T$ denote the sub-vector formed by the elements of $\x$ whose indices are listed in the set $\T$.
$\T^c$ denotes the complement of the set $\T$ w.r.t. the set $\{1, 2, \dots, N\}$. For a set $\T$, $|\T|$ denotes its cardinality (size),
and for a scalar c,  $|c|$ denotes the magnitude of $c$. $\A^T$, and $\A^\dagger$ denote the transpose, and pseudo inverse
of the matrix $\A$ respectively. $\mathbb E$ denotes the expectation operator which is approximated by the sample mean taken over a large
number of trials.}}
Compressed Sensing (CS) \cite{Donoho2006Compressedsensing,Candes2006RobustUncertainty} uses sparsity of the signal to reduce the number of
measurements required to represent the
signal. In general, reconstruction of the signal from the compressed measurements is NP-hard. In literature, a variety of suboptimal
solutions in polynomial time have been proposed for this purpose. They can be broadly classified as Convex Relaxation
methods \cite{Candes2005Decoding, Scott1998AtomicBP}, Bayesian framework
\cite{Shihajo2008BayesianCompressiveSensing,Wipf2004SparseBayesianLearning}, and
Greedy Pursuits (GP)
\cite{Mallat1993MatchingPursuits,Tropp2007SigRecoveryOMP,WeiDai2009SubspacePursuit}.
\par 
We consider only GP in this paper. GP iteratively estimate the support-set by selecting one or more atoms in each iteration,
until a convergence criteria is
met. In general, each iteration of GP consists of two steps: atom(s) selection step , and residual updating step. Popular examples
of GP include Matching Pursuit (MP) \cite{Mallat1993MatchingPursuits}, Orthogonal Matching Pursuit (OMP)
\cite{Tropp2007SigRecoveryOMP}, Subspace Pursuit (SP)\cite{WeiDai2009SubspacePursuit}, and Compressive Sampling Matching Pursuit (CoSaMP)
\cite{Needell2009CoSaMP}.
\par
Empirically, it has been observed that the recovery performance of the GP varies and depends on the nature of the sparse
signal \cite{Sturm2011AStudyonSparseVectorDistbns,Maleki2010OptimallyTuned}. For e.g., OMP may perform better than SP or {\em vice versa}
for some
types of signals. If the underlying statistical distribution of the non-zero values of the signal is known {\em a priori}, we can use
the best recovery algorithm suitable for that type of signal. But in many practical situations, we may not have this prior knowledge and
hence, we cannot use the best method as the {\em best} method is signal dependent. 
\par
It can be also seen that any greedy pursuit algorithm requires a minimum number of measurements, which is method dependent, for sparse signal
recovery. Also, all GP perform poorly in very low dimension measurement regimes. But often many applications provide very less number of
measurements and hence lower dimension measurement regimes are particularly important in reality.
\par
To address these issues, we propose a novel fusion
framework which fuses the information from two GP and estimates the
correct atoms from the union of the support-sets of the two which we refer to as {\em Fusion of Greedy Pursuits} (FuGP).
In this paper, we use OMP and SP as two examples of GP, but they can be replaced with any other GP.
\section{CS framework and Greedy Pursuits}
Consider the standard signal acquisition model which acquires a signal
$\mathbf x
\in \mathbb{R}^N$ via linear measurements using
\begin{equation}
\label{Eqn:CS}
\mathbf b = \mathbf  {Ax} + \mathbf w
\end{equation}
where $\mathbf A \in \mathbb{R}^{M\times N}$ represents a measurement (sensing) matrix, $\mathbf b \in \mathbb{R}^M$ represents the
measurement vector, and $\mathbf w \in \mathbb{R}^M$  represents the additive measurement noise in the system. In  CS framework, we have,
$M \ll N$ and (\ref{Eqn:CS}) is a well known ill-posed problem. But with the additional knowledge that the signal is $K$-sparse\footnote{A
signal is
said to be $K$-sparse if at most $K$ of its elements are non-zeros.} ($K < M$),
we can uniquely recover the signal under a few assumptions \cite{Candes2006RobustUncertainty, Donoho2006Compressedsensing}. 
\par
OMP \cite{Tropp2007SigRecoveryOMP} selects one prominent atom which gives the maximum correlation value between the columns of
$\mathbf A$ and the regularized measurement vector in every iteration. OMP finds support-set of a $K$-sparse signal in $K$
iterations. 
\par
SP \cite{WeiDai2009SubspacePursuit}  first selects the prominent $K$ columns of $\mathbf A$ from a matched filter output. In
subsequent
steps of the iterations, SP refines the initial solution by performing a test for subsets of $K$ columns in a group, and maintains a list of
$K$ columns of $\mathbf A$. The refinement of the solution set continues as long as the $l_2$-norm of the residue decreases. 

\section{Fusion Framework for GP }
We will start with an experiment which shows the motivation of the proposed fusion framework and its significance in sparse recovery.
\par
Consider a CS system where the signal dimension is 500, and the sparsity level is 20. Using our notations, we have, $N=500$ and $K=20$.
In this example, let us assume that the signal is a Gaussian sparse signal in a clean measurement setup (see Section \ref{Sec:Simulation}
for more details about the
simulation setup). We consider two CS sparse recovery algorithms viz. OMP and SP for reconstruction of the signal.
Let $\T$ denotes the actual support-set, and $\THat_{OMP}$ and $\THat_{SP}$ denote the support-sets estimated by OMP and SP
respectively. Let $\THat^{true}_{OMP} =  \T \cap \THat_{OMP}$ and $\THat^{true}_{SP} =  \T \cap \THat_{SP}$ represent the sets
of true atoms estimated by OMP and SP respectively. Then, we have, $|\T| = |\THat_{OMP}| = |\THat_{SP}| = K$, $0 \le
|\THat^{true}_{OMP}| \le K$, and $0 \le |\THat^{true}_{SP}| \le K$.
\par
Table \ref{Tab:motivationFuGP} presents the average (over $10,000$ trials) number of true atoms in the estimated support-sets for
Gaussian sparse signals for different values of $\alpha$ where $\alpha$ is defined as 
\begin{align}
\label{eqn:delta}
 \alpha = M/N.
\end{align}
$\alpha$ denote the fraction of measurements, also called the normalized measure of problem indeterminacy \cite{Maleki2010OptimallyTuned}.
The details of the simulation are given in Section \ref{subsec:Simulation}. 
It can be seen from Table \ref{Tab:motivationFuGP} that for $\alpha = 0.12$, the average number of correctly estimated atoms by OMP is
$8.1$, and
by SP is $10.5$. Interestingly, the average number of correct atoms in the union of the support-sets estimated by OMP and SP is 12.4, closer
to the true value 20. Also, using the property of union operator in set theory, it is guaranteed that the union of the estimated
support-sets always contain at least as many true atoms as in the estimated support-set of the best performing algorithm.

\begin{table}[!h]
\begin{center}
\begin{tabular}{|c|c|c|c|c|c|}
\hline
$\displaystyle \alpha = M/N$ & 0.10 & 0.11 & 0.12 & 0.13 & 0.14 \bigstrut \\ \hline
 $\displaystyle Avg |\hat \T^{true}_{OMP}|$ & 5.6 &     6.7 &     8.1 &    10.1 &    12.6 \bigstrut \\ \hline 
  $\displaystyle Avg|\hat \T^{true}_{SP}|$ &   5.8 &     7.9 &    10.5 &    13.2 &    15.6 \bigstrut \\ \hline 
   $\displaystyle Avg |\hat \T^{true}_{OMP} \cup \hat \T^{true}_{SP}|$  & 7.9 &     9.9 &    12.4 &    15 &    17.1  \bigstrut \\
\hline 
\end{tabular}
\end{center} 
\caption{Average number of correctly estimated atoms by OMP and SP, for Gaussian sparse signal, in clean measurement case, averaged over
$10,000$ trials ($N = 500$, $K = 20$). }
\label{Tab:motivationFuGP}
\end{table}
\par
These observations lead to the possibility of estimating more correct atoms from the union set than that individually estimated by OMP and
SP algorithms. The exhaustive search among the atoms in the union set is $2K \choose K$ ($40 \choose 20$ in our experiment) in the worst
case, which is significantly small as compared to the original where it is $N \choose K$ ($500 \choose 20$ in our experiment). But for
larger values of $K$, $2K \choose K$ is still very large. Hence, by employing some efficient scheme to select $K$ atoms from $2K$
atoms, we may be able to improve the number of correctly estimated atoms and improve the quality of the reconstructed sparse signal. 

\subsection{Proposed Fusion Framework}
To develop the fusion framework using OMP and SP as ingredient algorithms, let us  call the union of the estimated support-sets as {\em
joint support-set} and  denote by $\Gamma = \THat_{OMP} \cup \THat_{SP}$ .
Also, we call the intersection of the estimated support-sets as {\em common support-set} and denote by $\Lambda = \THat_{OMP} \cap
\THat_{SP}$. We have, $|\THat_{OMP}|$ = $|\THat_{SP}|$ = $K$,  $0 \le |\Lambda|
\le K$ and $K \le |\Gamma| \le 2K$. In the fusion framework, our task is to pick $K$ atoms from the joint support-set with
$|\Gamma|$ atoms. Assuming $M \ge 2K$, we propose a least-squares based method in FuGP for this purpose.
\par
Now, let us define two algorithmic functions which will be used in the proposed sparse recovery algorithms.
\begin{mydef} Let $\A \in \R^{M\times N}$, $\b \in \R^{M\times 1}$ and $K$ be the sparsity level. Also let $\T_{init}$ denote
the initial support-set with $|\T_{init}| < K$. Then we define the following algorithmic functions.
\begin{align*}
\hat\T_{OMP} &= OMP(\A, \b, K, \T_{init}) \\
\hat\T_{SP} &= SP(\A, \b, K, \T_{init})
 \end{align*}	
where $|\hat\T_{OMP}| = |\hat\T_{SP}| = K$.
\end{mydef}
The functions ``{\em OMP}'' and ``{\em SP}'' execute Algorithms \ref{Alg:OMP_InitSupport} and \ref{Alg:SP_InitSupport} respectively. Note
that by putting $\T_{init} = \varnothing$ in Algorithms \ref{Alg:OMP_InitSupport} and \ref{Alg:SP_InitSupport}, we get classic OMP and SP
respectively.
 
\begin{algorithm}
\caption{OMP with Initial Support}
\label{Alg:OMP_InitSupport}
{\bf Inputs:} $\mathbf{A}_{M \times N}$, $\mathbf{b}_{M \times 1}$, $K$, and $\T_{init}$
\begin{algorithmic}[1]
\Ensure $|\T_{init}| \le K-1$
\State $k = |\T_{init}|$;
\State $\r_k = \b - \A_{T_{init}} \A_{T_{init}}^\dagger \b$;
\State $\THat_k = \T_{init}$;
\Repeat
\State $k = k+1$;
\State $i_k$ = $\displaystyle {\mathop{\arg \max}_{i=1:N, i \notin \THat_{k-1}}}{\a_i^T\r_{k-1}}$;
\State $\THat_k = i_k \cup \THat_{k-1}$;
\State $\r_k = \b - \A_{\THat_k} \A_{\THat_k}^\dagger \b$;
\Until{($k \ge K$)};
\State $\THat = \THat_{k}$;
\end{algorithmic}
{\bf Outputs:} $\THat$, $\x_\THat = \A_\THat^\dagger \b$, and $\x_{\THat^c} = \0$.
\end{algorithm}

\begin{algorithm}
\caption{SP with Initial Support}
\label{Alg:SP_InitSupport}
{\bf Inputs:} $\mathbf{A}_{M \times N}$, $\mathbf{b}_{M \times 1}$, $K$, and $\T_{init}$
\begin{algorithmic}[1]
\Ensure $|\T_{init}| \le K-1$
\State $k = 0$;
\State $\r_k = \b - \A_{T_{init}} \A_{T_{init}}^\dagger \b$;
\State $\THat_k = \T_{init}$;
\Repeat
\State $k = k+1$;
\State $J$ = indices of the $K$ highest amplitude \\ \hspace{1.2cm}components  of ${{\mathbf A}^T} {\mathbf r}_{k-1}$;
\State $\tilde\T = J \cup \THat_{k-1}$;
\State $\v_{\tilde\T} = \A_{\tilde\T}^\dagger \b$, $\v_{\tilde\T^c} = \0$;
\State $\THat_k$ = indices corresponding to the $K$ largest \\ \hspace{1.2cm} magnitude
entries in $\v$; 
\State $\r_k = \b - \A_{\THat_k} \A_{\THat_k}^\dagger \b$;
\Until{($\vecLTwoNorm{\r_{k}} \ge \vecLTwoNorm{\r_{k-1}}$)};
\State $\THat = \THat_{k-1}$;
\end{algorithmic}
{\bf Outputs:} $\THat$, $\x_\THat = \A_\THat^\dagger \b$, and $\x_{\THat^c} = \0$.
\end{algorithm}

Since both methods (OMP and SP) agree on the atoms selected in $\Lambda$, we give more confidence on these atoms as compared to any other
atom in
$\Gamma$. Hence FuGP includes these atoms in the solution set. Now, we need to identify only $K-|\Lambda|$ atoms from the remaining
$|\Gamma|-|\Lambda|$ atoms.  Applying least-squares on the atoms in $\Gamma$, we form an intermediary solution for the signal.  The
remaining $K-|\Lambda|$ support atoms, we denote by $\tilde\T$, are
estimated as the $K-|\Lambda|$ indices  corresponding to the largest magnitudes in $\v$ which are not in $\Lambda$. Now, the support-set is
estimated as the union of the atoms in the sets
$\tilde \T$ and $\Lambda$. Finally, the sparse signal estimate is found from the estimated support-set using the least-squares.
The main steps of the FuGP algorithm for {\em fusing} estimated support-sets of OMP and SP are summarized in
Algorithm \ref{Alg:FuGP}.

\algrenewcommand{\algorithmiccomment}[1]{\hfill$\blacktriangleright$  #1}

\begin{center}
\begin{algorithm}
\caption{{\em FuGP} Algorithm}
\label{Alg:FuGP}
\begin{flushleft}
{\bf Inputs:} $\mathbf{A}_{M \times N}$, $\mathbf{b}_{M \times 1}$, and $K$
\end{flushleft}
\begin{algorithmic}[1]
\State $\THat_{OMP} = OMP(\A, \b, K, \varnothing)$; \Comment \small Using Algorithm \ref{Alg:OMP_InitSupport}\normalsize
\State $\THat_{SP} = SP(\A, \b, K, \varnothing)$;	\Comment \small Using Algorithm \ref{Alg:SP_InitSupport} \normalsize
\State \label{AlgStep:commonSupportSet} $\Lambda$ = $\THat_{OMP} \cap \THat_{SP}$; \Comment ($0 \le |\Lambda| \le K$)
\State \label{AlgStep:unionSupportSet} $\Gamma$ = $\THat_{OMP} \cup \THat_{SP}$; \Comment ($K \le |\Gamma| \le 2K$)
\State \label{AlgStep:proxyEstimte} $\v_\Gamma = \mathbf{A}_{\Gamma}^ \dag \mathbf{b}$, $\v_{\Gamma^c} = \0$;  \Comment \small intermediary
signal estimate \normalsize
\State \label{AlgStep:intermediarySignal} $\tilde\T$ = indices corresponding to the ($K-|\Lambda|$) largest \\ \hspace{0.6cm} magnitude
entries in $\v$ which
are not in $\Lambda$;
\State \label{AlgStep:estimatedSupportSet} $\THat$ = $ \tilde\T \cup \Lambda $;
\end{algorithmic}
{\bf Output:} $\THat$, $\x_\THat = \A_\THat^\dagger \b$, and $\x_{\THat^c} = \0$.
\end{algorithm}
\end{center}

\par
The computational demand of FuGP algorithm is a little more than the added computational cost of the two underlying methods. For e.g.,
the computational complexity of both OMP and SP are $\mathcal O(MNK)$ independently. The computational complexity of FuGP in this case
remains $\mathcal O(MNK)$. To save the running time, we can run both SP and OMP in parallel and then apply FuGP on the estimated support
sets. It may be also observed that the memory requirement of core part of FuGP (steps
\ref{AlgStep:commonSupportSet}-\ref{AlgStep:intermediarySignal} in Algorithm \ref{Alg:FuGP}) is only $\mathcal O(N)$.\\
{\em Remarks:}
\begin{itemize}
 \item The fusion framework and FuGP algorithm are scalable and can be easily extended to accommodate more than two greedy
pursuit algorithms.
\item The performance of FuGP directly depends on the number of correct atoms in the  joint support-set. Hence, we should choose the
underlying algorithms such that  join support-set contains maximum number of correct atoms.
\end{itemize}

\noindent{\underline{\em Iterative Fusion:}} Proposed FuGP aims to identify the correct atoms present the joint support-set. Hence it can at most
identify all the correct atoms in the joint support-set. But in lower dimensional measurement regimes, the joint support-set may not contain all
correct atoms and FuGP will surely miss the correct atoms which are not included in the joint support-set. To address this issue and hence
improve the performance, we propose an iterative version of the fusion algorithm called {\em Iterative
Fusion of Greedy Pursuits} (IFuGP). 

\par
 In $k^{\text {th}}$ iteration, we find the common support-set $\Lambda_k$ and call OMP and SP with $\Lambda_k$ as the
initial support-set to identify the remaining atoms in the support. $\Lambda_0$ is initialized as null set. $\Lambda_k$ is updated as the
common atoms in the newly estimated support-sets by OMP and SP. We fuse the support-sets newly estimated by OMP and SP to find the new
estimate of the support-set. The iteration continues as long as the $l_2$-norm of the residue decreases. This procedure opens a possibility
to include more correct atoms in the joint support-set $\Gamma_k$ which are not in the previous iteration $k-1$. It may observed that we
internally call FuGP in each iteration. 
\par
IFuGP algorithm is explained in Algorithm \ref{Alg:IFuGP}. Observe that Steps \ref{AlgStep:IFuGP_OMP}-\ref{AlgStep:IFuGP_support} in
Algorithm \ref{Alg:IFuGP} essentially forms the FuGP algorithm. IFuGP is computationally more demanding than FuGP.

\begin{algorithm}
\caption{{\em IFuGP} Algorithm}
\label{Alg:IFuGP}
{\bf Inputs:} $\mathbf{A}_{M \times N}$, $\mathbf{b}_{M \times 1}$, and $K$
\begin{algorithmic}[1]
\State $k = 0$;
\State $\r_0 = \b$;
\State $\Lambda_0 = \varnothing$;
\Repeat
\State $k = k+1$;
\State \label{AlgStep:IFuGP_OMP} $\THat_{OMP} = OMP(\A, \b, K, \Lambda_{k-1})$; \Comment \small Using Algorithm \ref{Alg:OMP_InitSupport}
\normalsize
\State $\THat_{SP} = SP(\A, \b, K, \Lambda_{k-1})$; \Comment \small Using Algorithm \ref{Alg:SP_InitSupport} \normalsize
\State $\Lambda_k$ = $\T_{OMP} \cap \T_{SP}$; \Comment \small $0 \le |\Lambda_k| \le K$ \normalsize
\State $\Gamma_k$ = $\T_{OMP} \cup \T_{SP}$; \Comment \small $K \le |\Gamma_k| \le 2K$ \normalsize
\State $\v_{\Gamma_k} = \mathbf{A}_{\Gamma_k}^ \dag \mathbf{b}$, $\v_{\Gamma_k^c} = \0$;  \Comment \small intermediary signal
estimate \normalsize
\State $\tilde\T_k$ = indices corresponding to the ($K-|\Lambda_k|$) largest \\ \hspace{0.6cm} magnitude
entries in $\v$ which are not in $\Lambda_k$;
\State \label{AlgStep:IFuGP_support} $\THat_k$ = $ \tilde\T_k \cup \Lambda_k $;	\Comment \small $|\THat_k| = K$\normalsize
\State $\r_k = \b - \A_{\THat_k} \A_{\THat_k}^\dagger \b$;
\Until{($\vecLTwoNorm{\r_{k}} \ge \vecLTwoNorm{\r_{k-1}}$)};
\State $\THat = \THat_{k-1}$;
\end{algorithmic}
{\bf Output:} $\THat$, $\x_\THat = \A_\THat^\dagger \b$, and $\x_{\THat^c} = \0$.
\end{algorithm}

\begin{figure*}[!t]
 \vspace{-0.6cm}
\subfloat[Clean Measurement]{\resizebox{3.1in}{2.5in}{\includegraphics{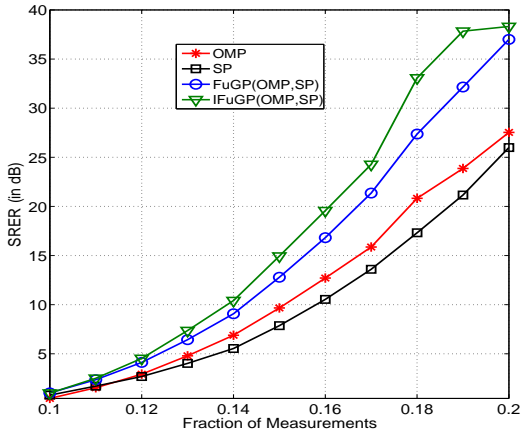}}}
\hfil
\subfloat[Noisy Measurement ({\em SMNR =15 dB})]{\resizebox{3.1in}{2.5in}{\includegraphics{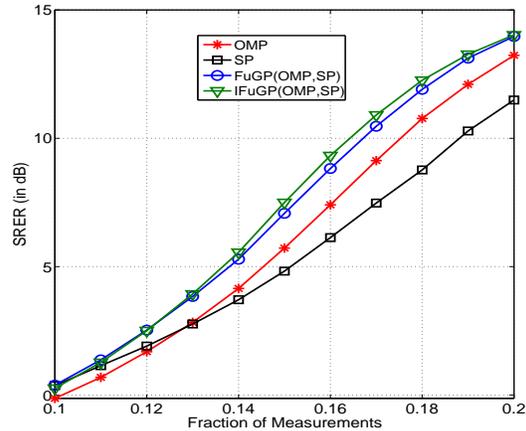}}}
\caption{Gaussian sparse signals: Signal-to-Reconstruction-Error Ratio (SRER) vs. Fraction of Measurements ($N = 500$, $K = 20$)} 
\label{Fig:Gauss}
\vspace{-0.3cm}
\end{figure*}
\par
In this paper, for notational brevity, we denote FuGP(OMP,SP) and IFuGP(OMP,SP) by FuGP and IFuGP respectively.

\section{Simulation and Results}
\label{Sec:Simulation}
We did extensive Monte Carlo simulations to evaluate performance of the proposed methods. We explain the simulation setup and define the
performance measure used for comparing different methods, in this section.
\subsection{Performance Measure} 
\subsubsection*{Signal-to-Reconstruction-Error Ratio (SRER)} Let $\x$ and $\hat\x$ denote the original and reconstructed sparse signal
vector. SRER is a performance measure build on top of Mean Square Error information. SRER (in dB) is defined as
\begin{equation}
\label{eqn:SRER}
 \textit{SRER (in dB)} \triangleq 10 \log_{10}\frac{\E\|\x\|_2^2}{\E\|\x-\hat\x\|_2^2}
\end{equation}

\subsubsection*{Signal to Measurement-Noise Ratio (SMNR):} Let $\sigma_s^2$ and $\sigma_n^2$ denote the power of each
element of signal and noise vector respectively. For noisy measurement simulations, we define SMNR in $dB$ as 
\begin{equation}
\label{Eqn:SMNR}
 \textit{SMNR (in dB)} = 10 \log_{10} \frac{\mathbb E \|\mathbf x\|_2^2}{\mathbb E \|\mathbf w\|_2^2} = 10 \log_{10} \frac{K
\sigma_s^2}{M\sigma_n^2}.
\end{equation}
\subsection{Experimental Setup}
\label{subsec:Simulation}
 Many of the GP including OMP and SP provide theoretical guarantees for convergence, but the theoretical bounds are by
and large ``pessimistic'' worst case bounds. In general, all CS sparse recovery methods work efficiently near this region.   But in many
applications, the number of measurements may be very limited due to many practical
reasons. Hence we are more interested in the lower dimensional measurement regimes where the sparse recovery methods begin to collapse. 
To compare the performance of the proposed methods with OMP and SP in this highly under-sampled region, we choose small values of
$\alpha$ where $\alpha$ is defined in \eqref{eqn:delta}.\\
The main steps involved in the simulation are the following:
\begin{enumerate}
\item Fix $K$, $N$ and choose an $\alpha$ so that the number of measurements $M$ is an integer.
\item \label{step2}Generate elements of $\mathbf A_{M \times N}$ independently from $\mathcal N(0,\frac{1}{M})$ and normalize each column
norm to unity.
\item \label{step3} Choose $K$ locations uniformly over the set \{1,2\dots N\} and fill these locations of $\mathbf x$ based on the choice
of signal characteristics:
\begin{enumerate}
 \item {\em Gaussian sparse signal:} non-zero values independently from $\mathcal N(0,1)$.
 \item {\em Rademacher sparse signal:} non-zero values are set to +1 or -1 with probability $\frac{1}{2}$. They are also known as
``constant amplitude random sign'' signals. 
\end{enumerate}
Set remaining $N-K$ locations of $\mathbf x$ as zeros.
\item For noisy regime, the additive noise ${\mathbf w}$ is a Gaussian random vector whose elements are independently chosen from $\mathcal
N(0,\sigma_w^2)$ and for clean case, $\mathbf w$ is set to zero.
\item The measurement vector  $\mathbf b = \mathbf A \mathbf x + \mathbf w$.
\item \label{step5} Apply the reconstruction methods independently.
\item \label{step6} Repeat steps \ref{step3}-\ref{step5} $T$ times. $T$ indicates the number of times $\mathbf x$ independently
generated, for a fixed ${\mathbf A}$.
\item Repeat steps \ref{step2}-\ref{step6} $S$ times. $S$ indicates the number of times $\mathbf A$ is independently generated.
\item \label{step8} Calculate SRER by averaging over $S \times T$ data 
\item Repeat steps \ref{step2}-\ref{step8}  for different values of $\alpha$.
\end{enumerate}

\begin{figure*}[!t]
 \vspace{-0.6cm}
\subfloat[Clean Measurement]{\resizebox{3.1in}{2.5in}{\includegraphics{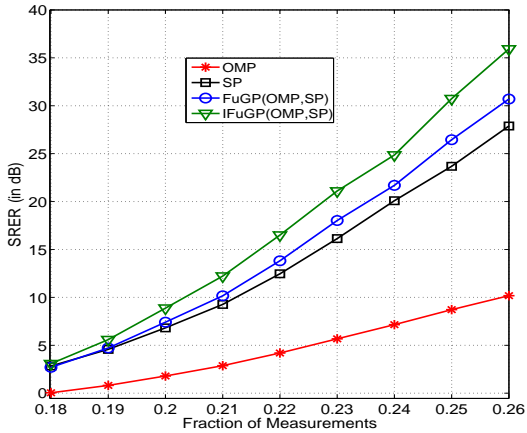}}}
\hfil
\subfloat[Noisy Measurement ({\em SMNR =
15 dB})]{\resizebox{3.1in}{2.5in}{\includegraphics{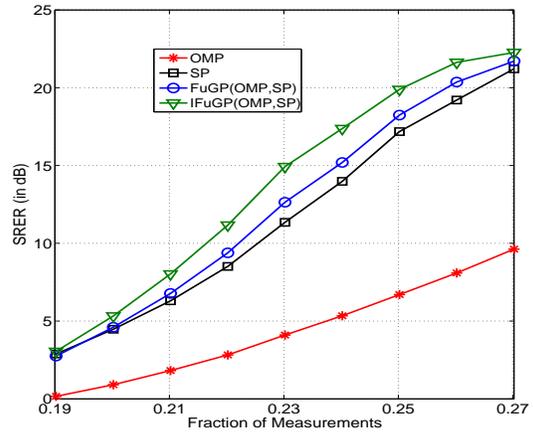}}}
\caption{Rademacher sparse signals: Signal-to-Reconstruction-Error Ratio (SRER) vs. Fraction of Measurements ($N = 500$, $K = 20$)} 
\label{Fig:Bernoulli}
\vspace{-0.3cm}
\end{figure*}

\subsection{Results and Discussions}
We performed Monte Carlo simulation with following parameters. $N=500$, $K=20$, $S=100$, $T=100$. i.e., we generated the
measurement matrix $\A$ 100 times and for each realization of $\A$, we generated a sparse signal with
ambient dimension 500 and sparsity level $K=20$, 100 times. 
\subsubsection*{Gaussian Sparse Signal} The performance of FuGP and IFuGP with OMP and SP as ingredient methods for Gaussian sparse signals
in clean as well as noisy measurement cases are shown in Fig.\ \ref{Fig:Gauss}. FuGP showed a significant improvement as compared to the
ingredient methods OMP and SP in both cases. IFuGP was able to give an improvement over FuGP. For example, in Fig.\ \ref{Fig:Gauss}(a), for
$\alpha = 0.18$, FuGP gave 6.5 dB ($31\%$) and 10 dB ($58\%$) SRER improvement respectively over OMP and SP in clean measurement case. For the same
scenario, IFuGP(OMP, SP) was able to improve the performance further and showed 12 dB ($59\%$), and 16 dB ($91\%$) improvement respectively over  OMP and SP. Compared to FuGP, IFuGP gave 5.72 dB ($21\%$) additional performance improvement in SRER.
\par
In the noisy measurement case (refer Fig.\ \ref{Fig:Gauss}(b)), for $\alpha = 0.18$, FuGP improved the performance by 1.1 dB ($11\%$) and
3.1 dB ($36\%$) over OMP and SP respectively. For the same situation, IFuGP gave 1.5 dB ($13\%$), and 3.5 dB ($40\%$) additional SRER as compared to OMP and SP
respectively, and also showed 0.35 dB ($3\%$) SRER improvement over FuGP.
\subsubsection*{Rademacher Sparse Signal} The results of simulation for Rademacher sparse signal is shown in Fig.\ \ref{Fig:Bernoulli}. Here
also FuGP and IFuGP showed performance improvement over OMP and SP. In the clean measurement case (refer Fig.\
\ref{Fig:Bernoulli}(a)), for $\alpha = 0.25$, FuGP gave 18 dB ($203\%$) and 2.8 dB ($12\%$) SRER improvement that OMP and SP respectively. IFuGP further
improved the performance and gave SRER improvement by 22dB ($252\%$) and 7 dB ($30\%$) respectively over OMP and SP. In this case IFuGP showed 4.3 dB ($16\%$)
SRER improvement over FuGP.
\par
In the noisy measurement simulation (refer Fig.\ \ref{Fig:Bernoulli}(b)), for $\alpha = 0.25$, by employing FuGP, an additional SRER
improvement of 11.5 dB ($172\%$) and 1.1 dB ($6\%$) was achieved as compared to OMP and SP respectively. In this case also, IFuGP continued to improve the
performance over FuGP (1.7 dB ($9\%$) SRER improvement than FuGP) resulting in 13.2 dB ($196\%$) and 2.7 dB ($16\%$) SRER improvement over OMP and SP respectively. 
\par
From the simulation results, it can be seen that FuGP and IFuGP improved the sparse signal recovery consistently in all the cases as
compared to the ingredient methods. The robustness against noise was shown in noisy measurement simulations for an SMNR = $15$ dB, which
closely matches many application scenarios.
\subsubsection*{Reproducible Results} In the spirit of reproducible research, we provide necessary Matlab codes downloadable in the following website: {\tt \url{http://www.ece.iisc.ernet.in/~ssplab/Public/FuGP.tar.gz}}. The code reproduces the simulation results shown in Fig.\ \ref{Fig:Gauss}, and Fig.\ \ref{Fig:Bernoulli}.

\section{Conclusions}
We proposed a novel fusion framework for Greedy Pursuits and also proposed two algorithms to recover the sparse signals. Using
simulations we showed that the proposed scheme can improve the sparse signal recovery performance of Greedy Pursuits in clean as well as noisy measurement cases. 
\bibliography{myBibtex}{}
\bibliographystyle{IEEEbib}%
\end{document}